\documentclass[12pt]{article}
\usepackage{geometry}            
\usepackage{graphicx}
\usepackage{amssymb}
\usepackage{amsmath}
\usepackage{epstopdf}
\usepackage{comment}

\usepackage[hyperindex=true,
          pdfstartview=FitH,
          bookmarksnumbered=true,
          bookmarksopen=true,
          citecolor=blue,
          linkcolor=blue,
          colorlinks=true,
          unicode]{hyperref}

\allowdisplaybreaks

\begin{document}

\title{Hamiltonian description of singular Lagrangian systems with spontaneously broken\\ time translation symmetry
}
\author{Liu Zhao, \,Pengfei Yu\, 
and \,Wei Xu\\
School of Physics, Nankai university, Tianjin 300071, China\\
{\em email}: 
\href{mailto:lzhao@nankai.edu.cn}{lzhao@nankai.edu.cn},
\href{mailto:yupengfei@foxmail.com}{yupengfei@foxmail.com}
\\
and \href{mailto:xuweifuture@mail.nankai.edu.cn}{xuweifuture@mail.nankai.edu.cn}}
\date{}                             
\maketitle

\begin{abstract}
Shapere and Wilczek recently found some singular Lagrangian systems which 
spontaneously breaks time translation symmetry. The common feature of their models 
is that the energy functions are multivalued in terms of the canonical phase space 
variables and the symmetry breaking ground states are all located at the brunching point 
singularities. By enlarging the phase space and making use of Dirac's theory on 
constrained Hamiltonian systems, we present the Hamiltonian description of some of  
the models discussed by Shapere and Wilczek and found that both the multivaluedness 
and the brunching point singularities can be avoided, while the spontaneous breaking of
time translation becomes more transparent. It is also shown that the breaking of time translation is always accompanied by the breaking of time reversal.

\vspace{3mm}
\vspace{3mm}


\end{abstract}

\section{Introduction}

Conventional wisdom says that every classical conservative mechanical system has a conserved total energy $H$ which generates a continuous time translation symmetry. 
Until very recently, it is commonly believed that time translation symmetry cannot be 
spontaneously broken. This picture has 
changed drastically due to the work \cite{Shapere:2012p1818} 
of Shapere and Wilczek, who found that 
in some special Lagrangian systems the time translation symmetry can be spontaneously 
broken in the lowest energy state (or called ground state). All models considered by 
Shapere and Wilczek bears the same property that the energy functions cannot be made 
singlevalued in terms of the generalized coordinates and their conjugate momenta. 
Historically, such systems are called singular Lagrangian systems 
\cite{Anonymous:2012p1906,Anonymous:2012p1907} or 
multivalued Hamiltonian systems \cite{Anonymous:2012p1901}, and, even though a 
plausible quantization procedure exists \cite{Anonymous:2012p1901} in certain special 
cases, a systematic Hamiltonian description for the 
classical case is still missing.

In this paper, we shall address the Hamiltonian description of singular Lagrangian systems with spontaneously broken time translation symmetry. Before doing this, we 
would like to pay some words clarifying the meaning of spontaneous breaking of time 
translation symmetry. First of all, in Hamiltonian mechanics, states of motion are 
represented by points in the phase space. With properly chosen coordinates on the 
phase space, we can say that a state of motion is just a point in phase space with definite
coordinate values. The physical notion of motion is just the change of the state in the
course of time governed by the classical equations of motion, 
and the ground state is the state that minimizes the Hamiltonian function,  
usually subjects to no motion. 
For standard canonical Hamiltonian systems, 
the most commonly used coordinates on the phase space are 
generalized coordinates and their conjugate momenta, which constitute a set of Darboux
coordinates on the phase space regarded as a symplectic manifold. However, this does 
not mean that Darboux coordinates are the only viable coordinates on the phase space.  
In principle, every set of $2n$ ($n$ being the number of degrees of freedom of the 
mechanical system) linearly independent functions of the Darboux coordinates can be 
chosen as a coordinate system on the phase space. Let $\{X_{i}, i=1\cdots 2n\}$ be 
any coordinate system on the phase space, under which  the ground state 
has coordinates $\{X^{G}_{i}\}$. A state is subject to nontrivial motion if there 
exist some $i$ such that $\dot X_{i}\equiv \frac{d}{dt}X_{i}\ne 0$. Recall that under 
infinitesimal time translation $t\rightarrow t+\epsilon$, $X_{i}(t)\rightarrow 
X_{i}(t+\epsilon)\simeq X_{i}(t)+\epsilon\dot X_{i}(t)$.  If $\dot X^{G}_{i}=0$ for all 
$i$, then the ground state is preserved by time 
translation symmetry. This is the case for most familiar mechanical systems (e.g. classical 
harmonic oscillator). If, on the contrary, $\dot X^{G}_{i}\neq 0$ for some $i$, then 
the ground state 
is not preserved by time translation, i.e. the time translation symmetry is 
spontaneously broken. We see that spontaneous breaking of time translation 
symmetry requires nontrivial motion in the ground state.

One may wonder how a true ground state of classical conservative mechanical system can carry through nontrivial motion. This question is beautifully answered by Shapere 
and Wilczek by presenting concrete examples. The models they discussed are all 
Lagrangian systems depending on higher powers of generalized velocities. 
For such systems
the conjugate momenta are nonlinear functions in terms of generalized velocities, and 
this makes the Hamiltonian multivalued in the conjugate momenta. The trick is that the 
ground states always appear at the branching points (they called these the ``turning 
points'') 
so that their motion is simultaneously governed by several different Hamiltonian 
functions, rendering the result of the motion unpredictable. From the point of view of 
Hamiltonian mechanics, the multivaluedness of the Hamiltonian signifies that the 
system is ill defined, and the ground states at the branching points of the multivalued 
Hamiltonian are in contradiction with the usual smooth minimization procedure. Below 
we shall present Hamiltonian descriptions of some of the models considered by Shapere 
and Wilczek, avoiding the appearance of multivalued Hamiltonian and the branching 
point singularities. The only price needed to be paid is that we have to abandon the use 
of Darboux coordinates on the phase space.

\section{General considerations}

Consider the following Lagrangian involving higher powers of generalized velocity,
\begin{align}
\mathcal{L}=\sum_{k=1}^{n}\frac{1}{2k}f_{k}(\phi)\dot \phi^{2k}-f_{0}(\phi), 
\qquad(n>1)  \label{l}
\end{align}
where $\phi$ denotes the generalized coordinate on the configuration space and $f_{k}
(\phi)$ are functions of $\phi$ with 
$f_{n}(\phi)> 0$ and at least one of the $f_{k}(\phi)$ is negative for $k=1,...,n-1$. This 
Lagrangian contains only even powers of $\dot\phi$ because we 
implicitly imposed a time reversal symmetry. 

The conjugate momenta associated with $\phi$ reads
\[
p=\frac{\delta\mathcal{L}}{\delta\dot\phi}=\sum_{k=1}^{n}f_{k}(\phi)\dot
\phi^{2k-1}.
\]
Following the standard procedure, we introduce the Hamiltonian
\[
\mathcal{H}=p\dot\phi-\mathcal{L}
=\sum_{k=1}^{n}\frac{2k-1}{2k}f_{k}(\phi)\dot\phi^{2k}+f_{0}(\phi).
\]
The Hamiltonian should in principle be regarded as a function of the phase space 
variables $(\phi,p)$. However, since $p$ is a nonlinear 
polynomial in $\dot\phi$, $\mathcal{H}$ will in general be multivalued in 
$p$. So, for a given set of values of $(\phi,p)$, there can be several 
different values for $\mathcal{H}$. If, as the states tend to some special points 
$(\phi^{(B)},p^{(B)})$,
some of the different values of $\mathcal{H}$ become degenerate, we call these special 
points the branching points of the Hamiltonian.

The existence of branching points indicates that the original Lagrangian system is 
singular and possesses some special features. Indeed, though at the first glance, the 
Lagrangian depends only on $(\phi, \dot\phi)$ and the the standard Euler-Lagrangian 
equation of motion
\[
\frac{\delta\mathcal{L}}{\delta\phi}-\frac{d}{dt}\left(\frac{\delta\mathcal{L}}
{\delta\dot\phi}\right)=0
\]
should hold without problem, a moment thinking would reveal that things are more 
involved. Consider the action $S=\int \mathcal{L}dt$ for the above system. After 
integration by parts, the Lagrangian can be changed into a form containing higher order 
time derivatives of $\phi$. In the conventional treatment of such systems, one takes $
\phi,\dot\phi,...,\frac{d^{\ell-1}}{dt^{\ell-1}}\phi$ as independent generalized 
coordinates and define variational 
derivatives of the corresponding Lagrangian with respect to the time derivatives of these 
coordinates, i.e. $\dot \phi,\ddot\phi,...,\frac{d^{\ell}}{dt^{\ell}}\phi$, as generalized 
momenta ($\ell$ is the highest order of time derivatives appearing in the Lagrangian) . In 
doing so, one 
effectively turns the system into a constrained Lagrangian system with a larger number of  
degrees of freedom, and to get the consistent time evolution of the 
unconstrained degrees 
of freedom, Dirac's theory for constrained Hamiltonian systems 
\cite{Dirac} is often employed.

Actually, the idea of turning the Lagrangian system into a constrained Hamiltonian system is the key to resolve the problem of presenting a consistent Hamiltonian 
description for singular Lagrangian systems with spontaneously broken time translation 
symmetry. To employ Dirac's method, it is not necessary to carry out the integration by 
parts in the action. Rather, we can manually introduce novel coordinates
$\gamma, \rho$ and introduce a 
new Lagrangian
\begin{align}
\mathcal{\bar L}=\sum_{k=1}^{n}\frac{1}{2k}f_{k}(\phi)\rho^{2k}-f_{0}(\phi) 
+\gamma(\rho-\dot\phi) \label{lb}
\end{align}
which is equivalent to the original one by solving the $\gamma$ equation of motion and 
substituting back in $\mathcal{\bar L}$. Since the Lagrangian (\ref{lb}) depends only 
linearly on $\dot\phi$ and does not depend on $\dot\gamma, \dot\rho$, it is clear that 
this system is a constrained system with a larger phase space containing the phase space 
of the original system as a subspace. Our task is then turned into finding
the consistent Hamiltonian structure on an unconstrained subspace of the enlarged 
phase space which gives the proper Hamiltonian description of our problem at hand. 
This last step, however, must be carried out on a case by case basis.

Let us remark that although in the above we considered only  systems involving 
a single generalized coordinate, the generalization to cases with multiple generalized 
coordinates is straightforward.

\section{Model analysis}

To make our analysis clear and understandable, we shall consider some of the models presented by Shapere and Wilczek, i.e. the $\dot\phi^{4}$ model, the $fgh$ 
model and the double sombrero model. These models will be analyzed in separate 
subsections.

\subsection{The $\dot\phi^{4}$ model}

The $\dot\phi^{4}$ model is unambiguously defined by its Lagrangian
\[
L(\phi)=-\frac{\kappa}{2}\dot\phi^{2}
+\frac{\lambda}{4}\dot\phi^{4},
\]
where $\lambda,\kappa>0$ are constant parameters. Clearly, this is the simplest case of 
$n=2$ in (\ref{l}) with the choice $f_{2}=\lambda, f_{1}=-\kappa$ and $f_{0}=0$. 
Using the idea outlined in the last section, we introduce the following Lagrangian
involving two novel coordinates $\rho$ and $\gamma$, 
\begin{align*}
\bar L(\phi, \rho,\gamma)=-\frac{\kappa \rho^2}{2}+\frac{\lambda}{4}\rho^4+
\gamma (\rho - \dot\phi).
\end{align*}
This system possesses an enlarged phase space spanned by $(\rho,\gamma,\phi,
\pi_{\rho},\pi_{\gamma},\pi_{\phi})$, where $(\pi_{\rho},\pi_{\gamma},
\pi_{\phi})$ are conjugate momenta of  the respective generalized coordinates. This 
enlarged, 6-dimensional phase space does not possess a well defined symplectic 
structure, because the definition of the conjugate momenta gives rise to 3 primary 
constraints,
\begin{align*}
G_{1}&= \pi_\phi+\gamma\simeq 0,\\
G_{2}&= \pi_\gamma\simeq 0,\\
G_{3}&= \pi_\rho\simeq 0.
\end{align*}
Using standard Legendre transformation we can obtain the expression for the 
Hamiltonian $H$ associated with $\bar L(\phi, \rho,\gamma)$. Consequently, we are 
ready to analyze the time evolution of the primary constraints with respect to the {\em 
total Hamiltonian}
\begin{align}
H_{total}&\equiv H+\sum_{i=1}^{3}\mu_{i}G_{i} \nonumber\\
&=\frac{\kappa }{2}{\rho}^2-\frac{\lambda}{4}{\rho}^4-\gamma\rho+\tilde
\mu_1G_{1}+\tilde\mu_2 G_{2}+\tilde\mu_3G_{3} \label{H1}
\end{align}
using the canonical Poisson brackets for the phase space variables $(\rho,\gamma,\phi,
\pi_{\rho},\pi_{\gamma},\pi_{\phi})$. It turns out that only the time evolution of  
$G_{3}$ leads to a secondary constraints
\[
G_{4}=-\kappa \rho+\lambda \rho^{3}+\gamma\simeq 0,
\]
and there is no further constraints. Notice that the Lagrangian multipliers $\mu_{i}$ 
and $\tilde\mu_{i}$ appearing on the first and second lines of (\ref{H1}) are not  
the same quantities. We have absorbed the unknown $\dot \phi, \dot \gamma, \dot 
\rho$ etc into the variables $\tilde\mu_{i}$.

Next we need to evaluate the matrix Poisson brackets between the constraints, i.e.
\[
M_{\alpha\beta}\equiv \{G_{\alpha}, G_{\beta}\}.
\]
The result reads
\[
M=\left(
\begin{matrix}
0 & 1 & 0& 0\cr
-1 & 0 & 0& -1\cr
0& 0& 0& \kappa-3\lambda\rho^{2} \cr
0 & -1& -\kappa+3\lambda\rho^{2} & 0
\end{matrix}
\right).
\]
As long as $\kappa-3\lambda\rho^{2}\neq 0$, the above matrix has rank 4 and all the constraints $G_{1}\sim G_{4}$ are of the second 
class. Following the standard procedure we introduce the Dirac Poisson 
bracket
\[
\{A, B\}_{DB}= \{A,B\} - \{A, G_{\alpha}\}(M^{-1})^{\alpha\beta}
\{G_{\beta},B\}
\]
for any two functions $A,B$ on the phase space. After doing so, all second class 
constraints can be consistently set equal to strong zeros.
The only remaining phase space variables are $(\phi,\rho)$, with
\begin{align}
\{\phi,\rho\}_{DB}=\frac{1}{3 \lambda {\rho}^2-\kappa},\quad 
\{\rho,\rho\}_{DB}=\{\phi,\phi\}_{DB}=0. \label{poissonbr}
\end{align}
The two dimensional subspace spanned by $(\phi,\rho)$ will be henceforth referred to as 
reduced phase space.
The final Hamiltonian on the reduced phase space is given by
\begin{align*}
H_{fin} = -\frac{1}{2} \kappa {\rho}^2+\frac{3}{4} \lambda {\rho}^4,
\end{align*}
and the Hamiltonian equations of motion are
\[
\dot \phi=\{\phi, H_{fin}\}_{DB}=\rho,\qquad \dot \rho=\{\rho,H_{fin}\}_{DB}
=0.
\]
The final Hamiltonian is smooth and single valued, its minima obey the usual
extremal conditions $\frac{\partial H_{fin}}{\partial \phi}=\frac{\partial H_{fin}}{\partial 
\rho}=0$. The minima occur at
\[
\rho^{G}=\pm\sqrt{\kappa/3\lambda},
\]
and $\phi^{G}$ is not fixed by the minimization condition. Rather, its rate of change in 
time is given by the equation of motion,
\[
\dot \phi^{G}=\rho^{G}=\pm\sqrt{\kappa/3\lambda}.
\]
If we take $(\phi^{G},\rho^{G})$ as parameters specifying the ground states, then the 
above system will possess two distinct families of infinitely degenerated ground states 
parametrized by $(\phi^{G},\rho^{G})=(\sqrt{\kappa/
3\lambda}t+\phi_{0},\sqrt{\kappa/3\lambda})$ and $(\phi^{G},\rho^{G})
=(-\sqrt{\kappa/3\lambda}t+\phi_{0},-\sqrt{\kappa/3\lambda})$ respectively, 
where the constant $\phi_{0}$ is the initial value of $\phi^{G}$. The explicit 
dependence of $\phi^{G}$ on $t$ indicates that the 
time translation symmetry is spontaneously broken. Moreover, the actual ground state 
must be chosen from only one of the two families of degenerate ground states. 
The concrete choice of ground state also breaks the time reversal 
symmetry.

What happens when $\kappa-3\lambda\rho^{2} = 0$? It appears at first glance that 
in this case the matrix $M$ becomes uninvertible and the above procedure for 
constructing Poisson brackets on the reduced phase space breaks down. This is also 
signified in the final Poisson bracket $\{\phi,\rho\}=\frac{1}{3\lambda\rho^{2}
-\kappa}$. If the singularity at the right hand side appears at other places in the phase 
space, this would indeed be a problem. However, notice that at exactly this particular 
singularity, the final Hamiltonian acquires its critical value. In the standard treatment of 
Hamiltonian mechanics, the symplectic structure on the phase space is often written as
\[
\Omega = dq \wedge dp = dt \wedge dH,
\]
where in the second equality the time parameter $t$ and the Hamiltonian $H$ are taken 
as a pair of action-angle variables. It is clear that the the critical points of $H$, the 
symplectic structure vanishes and the Poisson bracket is not well 
defined. Nonetheless, such singularities will not affect the effectiveness of the Poisson 
bracket in general, as we have witnessed above. In particular, when evaluating the 
equations of motion using the Poisson brackets (\ref{poissonbr}), all such singularities 
disappears and the resulting equations hold without problem even at the criticalities of 
the Hamiltonian.

\subsection{The $fgh$ model}

The previous model does not allow the inclusion of a potential, which makes it 
not useful when interaction is involved. To allow the inclusion of interacting potentials,  
let us look at a slightly more complex model -- the $fgh$ model specified by the 
Lagrangian
\[
L(\phi)=f(\phi){\dot\phi}^4+g(\phi){\dot\phi}^2+h(\phi)
\]
with $f(\phi)>0, g(\phi)<0$. This model is the special case of (\ref{l}) with 
$n=2$, $f_{2}=4f(\phi), f_{1}=2g(\phi)$ and $f_{0}=h(\phi)$.
Minimization of the corresponding energy function 
(here and below we shall omit the arguments of $f, g$ 
and $h$ and denote the derivatives such as $\frac{d f}{d\phi}$ as $f_{\phi}$ for 
simplicity)
\begin{align*}
E&=3f{\dot\phi}^4+g{\dot\phi}^2-h,\\
       &=3f\left({\dot\phi}^2+\frac{g}{6f}\right)^2
       -\frac{g^2}{12f}-h
\end{align*}
requires both ${\dot\phi}^2+\frac{g}{6f}=0$ and $\frac{g^2}
{12f}+h=c$, where $c$ is a constant. This is impossible for generic 
$h$, because the former condition implies nontrivial time dependence of $\phi$, while 
the latter requires $\phi$ to take some specific constant value(s). To circumvent this 
contradiction, it is necessary \cite{Shapere:2012p1818} to take $h$ 
to be the solution of the algebraic equation $
\frac{g^2}{12f}+h=c$. Doing so the Lagrangian is rewritten as
\[
L(\phi)=f{\dot\phi}^4+g{\dot\phi}^2+c-\frac{g^2}{12f}.
\]

We proceed as outlined in the last subsection and introduce an equivalent constrained Lagrangian,
\begin{align*}
\bar L(\phi, \rho,\gamma)=f\rho^4+g\rho^2+c-\frac{g^2}{12f}+\gamma (\rho - \dot
\phi).
\end{align*}
After a standard procedure for constructing Dirac Poisson bracket, we are left with a 
reduced phase space spanned by $(\rho,\phi)$ with the following Poisson
bracket and Hamiltonian,
\begin{align}
& \{\phi,\rho\}_{DB}=\frac{1}{2(6f\rho^{2}+g)}, \qquad
\{\phi,\phi\}=\{\rho,\rho\}=0,\\
&H_{fin} = 3f\rho^4+g\rho^2-c+\frac{g^2}{12f}
=3f\left(\rho^2+\frac{g}{6f}\right)^2-c.
\label{fgh}
\end{align}
The Hamiltonian equations of motion are
\begin{align*}
\dot \phi&=\{\phi, H_{fin}\}_{DB}=\rho,\\
\qquad \dot \rho&=\{\rho,H_{fin}\}_{DB}
=\frac{1}{24f^{2}}\left(f_{\phi}g-2g_{\phi}f\right)-\frac{1}{4f}f_{\phi}\rho^{2}.
\end{align*}
Once again, the apparent singularities in the $\{\phi,\rho\}_{DB}$ bracket appears 
at the critical points go $H_{fin}$, which is not a problem as we have explained in the 
last subsection.

Eq.(\ref{fgh}) shows that the final Hamiltonian is smooth and single valued, and its minima are not isolated points, but rather a pair of curves in the reduced phase space 
given by the equations $\rho=\pm\sqrt{\frac{-g}{6f}}$. It is easy to check that each of 
these two curves is consistent with the Hamiltonian equation of motion.

Due to the complicated form of the equations of motion, it is not easy to determine the time dependence of $\phi^{G},\rho^{G}$ in the ground states. However, given any 
value of $\phi^{G}$ consistent with the equation of motion, the states $
(\phi^{G},\rho^{G})=\left(\phi^{G},
\pm\sqrt{\frac{-g(\phi^{G})}{6f(\phi^{G})}}\right)$  
constitute two families of infinitely
degenerated ground states. The time dependence of each states in these 
two families indicates that the time translation symmetry is spontaneously broken.
Moreover, the actual ground state must be chosen from only one of the two families. 
The concrete choice of ground state automatically breaks the time reversal symmetry.

\subsection{The double sombrero model}

The double sombrero model is a little bit more complicated than the previous two models because more degrees of freedom are involved. The model Lagrangian is
\begin{align}
L(\rho,\phi)=\frac{1}{4}({\dot\rho}^2+\rho^2{\dot\phi}^2-\kappa)^2+\frac{\mu}
{2}\rho^2-\frac{\lambda}{4}\rho^4, \label{dso}
\end{align}
where $\kappa, \mu,\lambda$ are all positive constants. 
Following the above procedure, we first introduce the first order Lagrangian,
\begin{align*}
\bar L(\rho,\phi,\varphi,\psi,\gamma,\eta)=\frac{1}{4}(\varphi^2+\rho^2\psi^2
-\kappa)^2+\frac{\mu}{2}\rho^2-\frac{\lambda}{4}\rho^4+\gamma (\varphi - \dot
\rho)+\eta (\psi - \dot\phi).
\end{align*}
The construction of Dirac Poisson brackets in this model is much involved comparing to 
the case of the previous two models because the presence of more constraints. However, 
with the help of existing algorithm \cite{Gerdt:1999p2054}, it is not difficult to write 
down a computer algebra procedure using {\it Maple.
} 
It turns out that the reduced phase subspace is spanned by $(\rho,\phi,\varphi,\psi)$,
with the following Dirac Poisson brackets:
\begin{align*}
\{\rho,\phi\}_{DB}&=0,\\
\{\rho,\varphi\}_{DB}&=\frac{3\rho^2\psi^2+\varphi^2-\kappa}
{(3\rho^2\psi^2+3\varphi^2-\kappa)(\rho^2\psi^2+\varphi^2-\kappa)},\\
\{\rho,\psi\}_{DB}&=-\frac{2\psi\varphi}
{(3\rho^2\psi^2+3\varphi^2-\kappa)(\rho^2\psi^2+\varphi^2-\kappa)},\\
\{\phi,\varphi\}_{DB}&=-\frac{2\psi\varphi}
{(3\rho^2\psi^2+3\varphi^2-\kappa)(\rho^2\psi^2+\varphi^2-\kappa)},\\
\{\phi,\psi\}_{DB}&=\frac{3\rho^2\psi^2+\varphi^2-\kappa}
{(3\rho^2\psi^2+3\varphi^2-\kappa)(\rho^2\psi^2+\varphi^2-\kappa)\rho^2},\\
\{\varphi,\psi\}_{DB}&=\frac{2\psi}{\rho}
\frac{2\rho^2\psi^2+\varphi^2-\kappa}{(3\rho^2\psi^2+3\varphi^2-\kappa)
(\rho^2\psi^2+\varphi^2-\kappa)},
\end{align*}
and
\[
\{\rho,\rho\}_{DB}=\{\phi,\phi\}_{DB}=\{\varphi,\varphi\}_{DB}
=\{\psi,\psi\}_{DB}=0.
\]
The final Hamiltonian on the reduced phase space reads
\begin{align}
H_{fin} &=\frac{1}{12}\left(3\rho^2\psi^2+3\varphi^2-\kappa\right)^2+
\frac{1}{4\lambda}\left(\lambda\rho^2-\mu\right)^2-\frac{\kappa^2}{3}-
\frac{\mu^2}{4\lambda},
\label{sombrero}
\end{align}
which yields the Hamiltonian equations of motion
\begin{align}
\dot \rho&=\{\rho, H_{fin}\}_{DB}=\varphi,  \label{eee1}\\
\dot \phi&=\{\phi,H_{fin}\}_{DB}=\psi,  \label{e2}\\
\dot \varphi&=\{\varphi, H_{fin}\}_{DB}
=\rho\psi^2-\frac{\rho(\lambda\rho^2-\mu)(3\rho^2\psi^2+\varphi^2-\kappa)}
{(3\rho^2\psi^2+3\varphi^2-\kappa)(\rho^2\psi^2+\varphi^2-\kappa)},
\label{e3}\\
\dot \psi&=\{\psi, H_{fin}\}_{DB}=-\frac{2\psi\varphi}{\rho}+\frac{2\psi\varphi
\rho(\lambda\rho^2-\mu)}{(3\rho^2\psi^2+3\varphi^2-\kappa)(\rho^2\psi^2+
\varphi^2-\kappa)}. \label{e4}
\end{align}

Eq. (\ref{sombrero}) shows that the final Hamiltonian is smooth and single 
valued. The minima of the Hamiltonian are determined by the conditions
\begin{align}
&3\rho^2\psi^2+3\varphi^2-\kappa=0, \label{mmmm}\\
&\lambda\rho^2-\mu=0 \label{rho}
\end{align}
together with the equations of motion. Concretely, from (\ref{rho}), we can get
\[
\rho^{G}=\pm\sqrt{\frac{\mu}{\lambda}},
\]
which are constants. Therefore, from the equation of motion $\dot\rho=\varphi$, we get 
\[
\varphi^{G}=0.
\] 
Inserting these results into (\ref{mmmm}), we get
\begin{align*}
  \psi^{G}=\pm\sqrt{\frac{\lambda\kappa}{3\mu}},
\end{align*}
and further, from the equation $\dot\phi=\psi$, we get
\[
\phi^{G}=\pm \sqrt{\frac{\lambda\kappa}{3\mu}}\, t+\phi_{0},
\]
where $\phi_{0}$ is a constant.

Finally, the ground states are identified with the following curves,
\[
(\phi^{G},\rho^{G},\psi^{G},\varphi^{G})
=\left(\epsilon_{2}\sqrt{\frac{\lambda\kappa}{3\mu}}\, t+\phi_{0},
\epsilon_{1}\sqrt{\frac{\mu}{\lambda}},
\epsilon_{2}\sqrt{\frac{\lambda\kappa}{3\mu}}\,,0
\right),
\]
where $\epsilon_{1}=\pm 1,\epsilon_{2}=\pm 1$ and are independent of each other.
The number of independent choices of $(\epsilon_{1},\epsilon_{2})$ is 4. Therefore, 
there are totally 4 distinct families of infinitely degenerate ground states in this model, 
each family subjects to nontrivial motion, which spontaneously breaks time translation. 
The actual ground states must be chosen from one of the 4 families. Any such choice 
breaks time reversal symmetry.

{\it Remarks:}

1. As in the previous two models, the apparent singularities on the right hand sides of 
(\ref{e3}) and (\ref{e4}) are located exactly at the critical points of $H_{fin}$, so these
do not give rise to any problem as explained earlier.

2. The consistency of Hamiltonian equations of motion with the corresponding Euler-Lagrangian equation of motion for the $\dot\phi^{4}$ model and $fgh$ model can be 
easily checked. For the double sombrero model, the consistency between Hamiltonian 
equations of motion and the Euler-Lagrangian equations of motion seems more 
complicated to check. We leave this check in the appendix.

3. The Hamiltonian equations of motion for the double sombrero model, eqs. (\ref{e3})
and (\ref{e4}), seems to be singular at the ground states because the indeterminate 
$0/0$ appears on the right hand side. However, this is completely superficial. Actually, 
if we first collect a common denominator and then substituting in the ground state 
values of the phase space variables, the right hand sides of both (\ref{e3}) and (\ref{e4}) 
vanishes.

\section{Discussions}

Spontaneous breaking of time translation symmetry is an amazing discovery. The existence of such mechanical systems indicates that there is a hidden piece of classical 
mechanics about which our understanding is still very poor. Nonetheless, the importance 
of such models cannot be overestimated. On the one hand, these models provide explicit 
counter examples to the inverse of Noether's first theorem (which roughly says that any 
conserved charge generates a continuous symmetry). On the other hand, similar 
Largrangian systems appear widely in modern theoretical physics such as the $k$-
inflation models in cosmology \cite{ArmendarizPicon:1999p1875}, ghost condensation 
\cite{ArkaniHamed:2003p2066}, higher curvature and $f(R)$ gravity models etc.

The original treatment \cite{Shapere:2012p1818} of 
Shapere and Wilczek on the spontaneous breaking of time translation symmetry has 
depended on the use of standard Darboux coordinates on the phase space. 
Using such variables, the multivaluedness of energy functions and the branching point 
singularities seem to be inevitable. However, our work shows that both the 
multivaluedness of energy functions and the branching point singularities can be 
avoided, the only sacrifice to be paid is that non Darboux coordinates have to be adopted 
in parametrizing the phase space. The use of non Darboux coordinates also resolves
the puzzling jumping effects at the ground states described by Shapere and Wilczek. 
Actually, from the detailed model analysis presented in the last section, it is clear that
there is no such jumping at the ground states using the non Darboux coordinates. 
We believe that the motion with momentum jumping in the ground states is just a 
signature indicating that the momentum, as part of the Darboux coordinates, is not a 
good coordinate for parametrizing the phase space in such models. As a by-product, 
we also revealed that the time reversal symmetry is broken by the choice of 
ground states in such models.

Is it possible to extract Darboux coordinates from our treatment on the Hamiltonian description? The answer is yes, but the original problems will reappear in terms of these 
coordinates. To be more explicit, let 
us take the simplest $\dot\phi^{4}$ model as an example. It is easy to see that both
$(q_{1},p_{1})=(\kappa\rho-\lambda \rho^{3}, \phi)$ and $(q_{2},p_{2})=(\phi,
-\kappa\rho+\lambda\rho^{3})$ are viable choices of Darboux coordinates. 
Using either choices of Darboux coordinates, the Hamiltonian becomes multivalued, 
either in $q_{1}$ or in $p_{2}$. The multivaluedness of the Hamiltonian in terms 
of Darboux coordinates also indicates that these coordinates are not good 
coordinates on the phase space. On this point, please also notice that 
$p_{1}$ ($p_{2}$ reps.) is nonlinear in $q_{1}$ ($q_{2}$ reps.), 
therefore $(q_{1},p_{1})$ (or $(q_{2},p_{2})$) does 
not parametrize the cotangent bundle of the configuration space as apposed to the 
cases of canonical Hamiltonian systems.  

For usual Hamiltonian systems, passing from classical to the quantum description is straightforward, one only needs to replace the canonical Poisson bracket 
$\{q,p\}_{PB}=1$ by the canonical commutator $[\hat q, \hat p] =i h$. However, 
for the models discussed in the present work, the Poisson brackets between the ``good 
coordinates'' on the phase space are all nonlinear, hence the canonical quantization 
process does not work. The systematic quantization of such models remain an open 
problem\footnote{After the first version of this paper appeared in arXiv, 
we have noticed that Shapere and Wilczek (together with Xiong) have gone further on 
the issue of quantization \cite{Shapere:2012p2115} \cite{Shapere:2012p2280}.}. 
Perhaps the path integral quantization for the $\dot\phi^{4}$ model 
\cite{Anonymous:2012p1901} can provide some clue for the other models. 

Another related issue is whether the spontaneous breaking of time translation symmetry can be dynamical, i.e. can we construct models which initially preserve time translation
but eventually evolve into a phase with spontaneous breaking of time translation 
symmetry? A moment thinking seems to indicate that this is possible. Take the $fgh$ 
model for example. If we take $f(\phi)$ to be a positive constant, $g(\phi)$ to be 
a function such 
that $g(\phi)\ge 0$ for the initial value of $\phi$ but decreasing as $\phi$ evolves in 
time and eventually becomes negative, then the ground states will transit from the initial
symmetry preserving phase $(\phi^{G},\rho^{G})=(\phi_{0},0)$ to the symmetry 
breaking 
phase $(\phi^{G},\rho^{G})=\left(\phi^{G},
\pm\sqrt{\frac{-g(\phi^{G})}{6f(\phi^{G})}}
\right)$. We leave the detailed study of this phenomenon for later works.

\section*{Appendix}

In this appendix we check the consistency between Euler-Lagrangian equations and Hamiltonian equations of motion for the double sombrero model.

Variation of the Lagrangian (\ref{dso}) with respect to $\phi$ yields
\[
(2\rho\dot\rho\dot\phi+\rho^2\ddot\phi)(\dot\rho^2+\rho^2\dot\phi^2-\kappa)
+2\rho^2\dot\phi(\dot\rho\ddot\rho+\rho\dot\rho\dot\phi^2+\rho^2\dot\phi\ddot
\phi)=0.
\]
Similarly, variation of (\ref{dso}) with respect to $\rho$ gives
\[
\ddot\rho(\dot\rho^2+\rho^2\dot\phi^2-\kappa)+2\dot\rho(\dot\rho\ddot\rho+\rho
\dot\rho\dot\phi^2+\rho^2\dot\phi\ddot\phi)+[(\lambda\rho^2-\mu)\rho-\rho\dot
\phi^2(\dot\rho^2+\rho^2\dot\phi^2-\kappa)]=0.
\]
We rearrange these two equations in the form
\begin{align}
  &a_1\ddot\phi+b_1\ddot\rho+c_1=0, \label{lineq}\\
  &a_2\ddot\phi+b_2\ddot\rho+c_2=0, \label{lineqs}
\end{align}
where
\begin{align*}
  &a_1=\rho^2(3\rho^2\dot\phi^2+\dot\rho^2-\kappa), &&b_1=2\rho^2\dot\rho
  \dot\phi, &&c_1=2\rho\dot\rho\dot\phi(2\rho^2\dot\phi^2+\dot\rho^2-\kappa),
  \\
  &a_2=2\rho^2\dot\rho\dot\phi, &&b_2=3\dot\rho^2+\rho^2\dot\phi^2-
  \kappa, &&c_2=(\lambda\rho^2-\mu)\rho-\rho\dot\phi^2(\rho^2\dot\phi^2-\dot
  \rho^2-\kappa),
\end{align*}
and take  (\ref{lineq}) and (\ref{lineqs}) as a system of linear algebraic equations for $
\ddot\phi$ and $\ddot\rho$. The solution is easily found to be
\begin{align}
  &\ddot\phi=\frac{a_1c_2-a_2c_1}{a_2b_1-a_1b_2},  \label{eq11}\\
  &\ddot\rho=-\frac{b_1c_2-b_2c_1}{a_2b_1-a_1b_2}. \label{eq12}
\end{align}
Inserting the expressions for $a_{1},a_{2},b_{1},b_{2},c_{1},c_{2}$ and after some
algebraic simplifications, we find that (\ref{eq11}) and (\ref{eq12}) are exactly the same 
as the results which we can get by substituting (\ref{eee1}) and (\ref{e2}) into (\ref{e3}) 
and (\ref{e4}).

\section*{Acknowledgment} 

This work is supported by the National  Natural Science Foundation of 
China (NSFC) through grant No.10875059. L.Z. would like to thank the 
organizer and participants of ``The advanced workshop on Dark Energy 
and Fundamental Theory'' supported by the Special Fund for Theoretical 
Physics from the National Natural Science Foundation of China with grant 
no: 10947203 for discussions.

\bibliographystyle{utcaps}

\providecommand{\href}[2]{#2}\begingroup\raggedright\endgroup

\end{document}